\documentclass[11pt,nofootinbib,floatfix]{revtex4}
\usepackage{mathrsfs}
\usepackage{graphicx}
\usepackage{amsmath}
\usepackage{amsfonts}
\usepackage{amssymb}
\usepackage{color}

\usepackage{epsfig}
\usepackage{CJK}
\usepackage{graphicx}
\usepackage{epsfig}
\usepackage{eepic}
\usepackage{bbm}
\usepackage{dcolumn}
\usepackage{bm}
\usepackage{ulem}

\usepackage{tikz}
\usetikzlibrary{graphs}
\usepackage{minipage-marginpar}

\newcommand{\omits}[1]{}

\def\bc{\begin{center}}

\def\ec{\end{center}}
\def\be{\begin{eqnarray}}
\def\ee{\end{eqnarray}}

\definecolor{dyellow}{rgb}{1.,0.8,.0}
\definecolor{myblue}{rgb}{.1,.1,.7}
\definecolor{dcyan}{rgb}{.0,.6,.6}
\definecolor{cyan}{rgb}{0.4,1.0,1.0}
\definecolor{dmagenta}{rgb}{0.6,0.0,0.6}
\definecolor{brown}{rgb}{0.6,0.2,0.}
\definecolor{darkblue}{rgb}{.0,.0,0.5}
\definecolor{darkred}{rgb}{0.75,0.0,0.0}
\definecolor{orange}{rgb}{1.,.6,.0}
\definecolor{dorange}{rgb}{0.8,.4,.0}
\definecolor{green}{rgb}{0.0,1.0,0.0}
\definecolor{darkgreen}{rgb}{0.0,0.6,0.0}
\definecolor{purple}{rgb}{.4,.0,.4}
\definecolor{lightgrey}{rgb}{0.7, 0.7, 0.7}
\definecolor{grey}{rgb}{0.4, 0.4, 0.4}

\newcommand{\nc}{\newcommand}
\nc{\rnc}{\renewcommand} \nc{\ket}[1]{\left | \, #1 \right \rangle}
\nc{\bra}[1]{\left \langle #1 \, \right |}
\nc{\ua}{\uparrow} \nc{\da}{\downarrow}

\nc{\braket}[2]{\langle\, #1\,|\,#2\,\rangle}
\nc{\ev}[1]{\langle\, #1\,\rangle}
\nc{\half}{\frac{1}{2}}

\nc{\prj}{\mathcal{P}} \nc{\hilb}{\mathcal{H}}
\nc{\pth}{\mathcal{C}} \nc{\inprod}[2]{\braket{#1}{#2}}
\nc{\upket}{\ket{\uparrow}} \nc{\downket}{\ket{\downarrow}}
\nc{\upbra}{\bra{\uparrow}} \nc{\downbra}{\bra{\downarrow}}

\begin{document}


\title{Surface growth scheme for bulk reconstruction and $T\bar{T}$ deformation}

\author{Hao-Chun Liang$^1$} \email{lianghch3@mail2.sysu.edu.cn}
\author{Jia-Rui Sun$^{1}$} \email{sunjiarui@mail.sysu.edu.cn}
\author{Yuan Sun$^2$} \email{sunyuan@csu.edu.cn}

\affiliation{${}^1$School of Physics and Astronomy, Sun Yat-Sen University, Guangzhou 510275, China}
\affiliation{${}^2$Institute of Quantum Physics, School of Physics, Central South University, Changsha 418003, China}


\begin{abstract}
In this paper, we study the dynamical connection between the surface growth scheme and the conformal field theory with $T\bar{T}$ deformation. By utilizing the extended one-shot entanglement distillation tensor network, we find that the iterative growth, i.e. radial evolution of homogenous and isotropic bulk minimal surfaces in asymptotically anti-de Sitter (AdS) spacetime can be mapped to the $T\bar{T}$ operator flow driven by the deformation parameter. Our results show that the $T\bar{T}$ deformation can provide a dynamical mechanism for the surface growth in asymptotically AdS spacetime, which may shed light on reconstructing bulk gravitational dynamics from the surface growth scheme.
\end{abstract}


\maketitle
\tableofcontents

\section{Introduction}
The anti-de Sitter/Conformal field theory (AdS/CFT) correspondence establishes a profound connection between a gravitational system in an asymptotically AdS spacetime and a conformal field theory (CFT) on its boundary~\cite{Maldacena:1997re,Gubser:1998bc,Witten:1998qj}. As a duality between two theories, the AdS/CFT correspondence or more general gauge/gravity duality also indicates that the boundary quantum system should encode the full information of the bulk gravity. The program to construct bulk gravity from its dual boundary quantum field theory is called bulk reconstruction~\cite{Hamilton:2006az,Czech:2012bh,Headrick:2014cta,Dong:2016eik,Cotler:2017erl,Faulkner:2017vdd,Harlow:2018fse,Bao:2019bib}. Progresses in recent years have shown that entanglement entropy and its holographic dual, i.e., the holographic entanglement entropy~\cite{Ryu:2006bv,Ryu:2006ef,Hubeny:2007xt} play a very important role in realizing the bulk reconstruction. However, difficulties still remain in the previous approaches of entanglement wedge reconstruction. For example, how to find a more fine description for the subregion-subregion duality, how to give a unified framework to construct both the bulk matter fields and bulk geometry, and how to construct bulk gravitational dynamics from the dual boundary quantum fields.

In recent works~\cite{Lin:2020ufd,Yu:2020zwk}, a surface growth approach for bulk reconstruction in asymptotically AdS spacetime was proposed which can not only probe the fine structure of bulk entanglement, but also able to construct both the bulk geometry and matter fields from boundary quantum system. In the surface growth approach, a given boundary subregion is partitioned into many smaller adjacent segments, each generating a bulk minimal surface, namely, the Ryu-Takayanagi (RT) surface, then treating these RT surfaces as the new boundary of the bulk spacetime, new layer of bulk minimal surfaces can continue to grow on them, which probe deeper regions of the bulk spacetime. By repeating this process layer by layer, the bulk minimal surfaces can probe arbitrary regions in the entanglement wedge, similar to the bubble growing. In addition, it has been shown that the surface growth scheme can be explicitly realized by the multi-scale entanglement renormalization ansatz (MERA)-like tensor network~\cite{Bao:2018,Lin:2020ufd,miyaji:1503,Vidal:2008,vidal:1812,vidal:2007,hayden0204},  which naturally gives an emergence of bulk geometry and indicates a dynamical connection between the fine structure of entanglement and bulk spacetime. For example, the entanglement of purification which captures the fine structure in subregions of the entanglement wedge, can be naturally derived from the surface growth scheme~\cite{Lin:2020yzf} and also suggests a selection rule for surface growth in the bulk reconstruction in the AdS/BCFT correspondence~\cite{Fang:2024mwp}.

On the other hand, a parallel avenue of research explored the $T\bar{T}$ deformation of two dimensional CFTs, which deforms the ultraviolet (UV) dynamics of the theory~\cite{Zamolodchikov:2004ce,Smirnov:2016lqw,Cavaglia:2016oda}. The $T\bar{T}$ deformation has received a lot of attentions recently. For example, from the field theory side, the correlation functions~\cite{Cardy:2019qao,Kraus:2018xrn,He:2019vzf,He:2019ahx,He:2020udl,Cui:2023jrb} and R\'{e}nyi or entanglement entropy~\cite{Donnelly:2018bef,Chen:2018eqk,Sun:2019ijq,Lai:2025thy} of $T\bar{T}$-deformed two dimensional field theories have studied, and the geometric formulation of generalized root-$T\bar{T}$ deformation has been obtained~\cite{Babaei:2024hti}. From the holographic perspective, the $T\bar{T}$ deformation can alter the infrared (IR) geometry and result in a radial cutoff at the boundary of the asymptotically AdS$_3$ spacetime~\cite{Guica:2019nzm,McGough:2016lol,Kraus:2018xrn}, see, for example, recent reviews on $T\bar{T}$ deformation~\cite{Jiang:2019epa,He:2025ppz}.

The renormalization group (RG) flow of the $T\bar{T}$-deformed theory maps to bulk evolution along the radial direction, which is very similar to the process of surface growth. This similarity motivates us to ask an interesting question: Can the iterative growth of bulk minimal surfaces be linked to the operator driven evolution induced by $T\bar{T}$ deformation? In this work, we will unify these two perspectives by utilizing the tensor network realization of the surface growth scheme and demonstrate that it can be mapped to the $T\bar T$-deformed CFT. Explicitly, we generalized the one-shot entanglement distillation (OSED) tensor network to describe the general surface growth process in which bulk minimal surfaces grow layer by layer, and then connected the radial evolution of bulk minimal surfaces to the $T\bar T$ operator flow via the generalized OSED tensor network. Our analysis revealed that the deformation parameter $\mu$ governs the iterative growth of bulk minimal surfaces, while the dynamics of stress tensor emerges from the hierarchical entanglement structure.

The paper is organized as follows. Section~\ref{sec-review sg} briefly reviews the original surface growth scheme, then section~\ref{sec-gen-sg} extends the OSED tensor network to describe the general surface growth and analyzes its continuum limit. Section~\ref{sec-TTbar} establishes the explicit connection between the isotropic surface growth and the $T\bar T$ deformation. Finally, section~\ref{sect:conclusion} gives the conclusions and discusses future related problems to be investigated.

\section{Review of the surface growth scheme}\label{sec-review sg}
The surface growth scheme provides an efficient and systematic framework for reconstructing bulk spacetime through hierarchical entanglement structures~\cite{Lin:2020ufd,Yu:2020zwk}. It has been shown that this scheme can be described via generalized form of the one-shot entanglement distillation (OSED) tensor network~\cite{Bao:2018}, which maps the iterative growth of bulk minimal surfaces to tensor network architectures.

\begin{figure}[htbp]
	\begin{center}
		\includegraphics[height=7cm,clip]{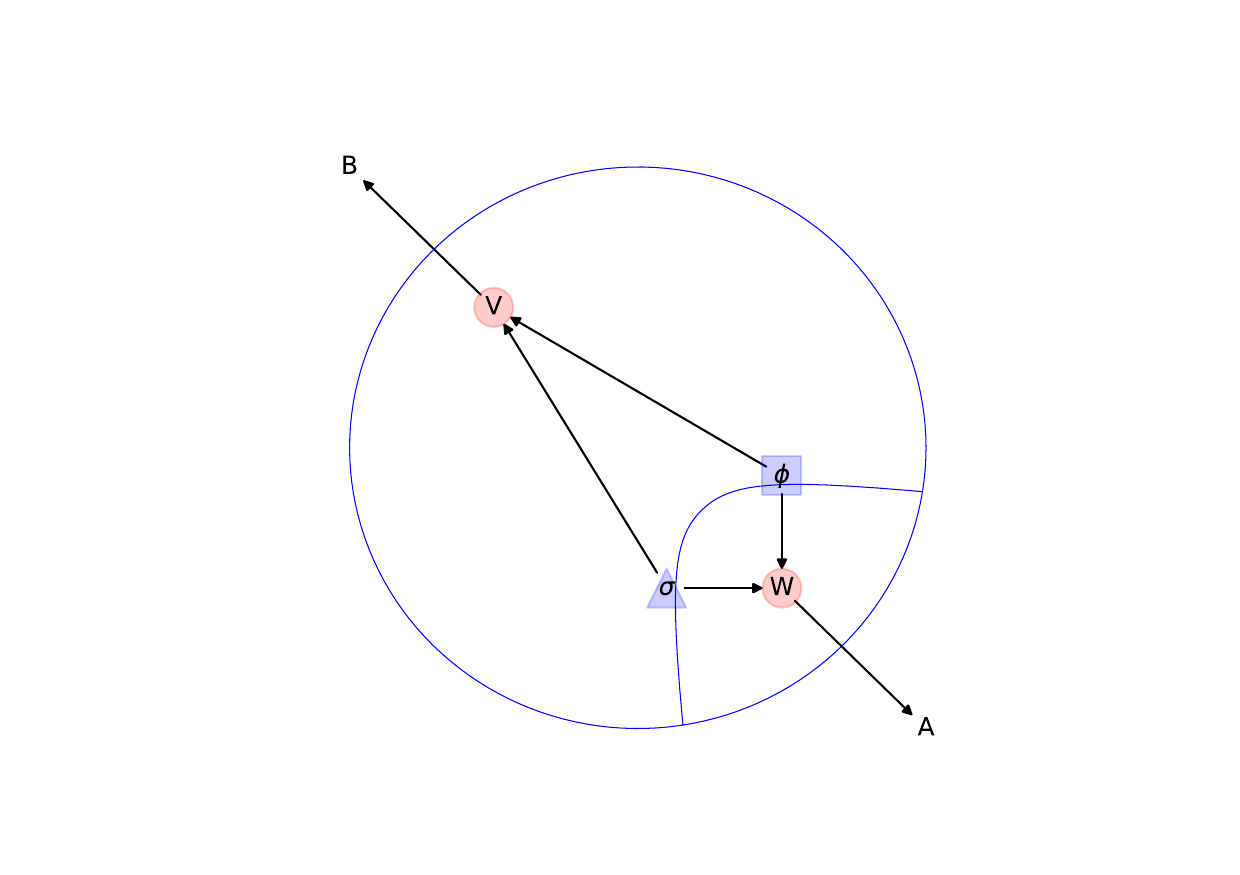}
		\caption{The bulk minimal surface $\phi\otimes\sigma$ grows out from the subsystem $A$. The isometry tensor $W$ matches to the entanglement wedge of $A$, and the isometry tensor $V$ matches to the rest of bulk spacetime.}
		\label{fig_original_AB}
	\end{center}	
\end{figure}

To illustrate, consider partitioning the spatial boundary of a CFT$_2$ into adjacent subsystems $A$ and $B$. The entanglement structure of $A$ is encoded through an isometric tensor network, as depicted in Fig.\ref{fig_original_AB}. The combined boundary state $\Psi^{AB}$ is expressed as:
\begin{equation}
\Psi^{AB}=V^B_{\alpha\beta} W^A_{\bar{\alpha}\bar{\beta}}\phi^{\alpha\bar{\alpha}}\sigma^{\beta\bar{\beta}},
\end{equation}
or equivalently,
\begin{equation}
\ket{\Psi}=(V\otimes W)(\ket{\phi}\otimes\ket{\sigma}),
\end{equation}
where $W$ and $V$ are isometric tensors corresponding to the entanglement wedges of $A$ and $B$, respectively. The states $\ket{\phi}$ and $\ket{\sigma}$ represent maximally entangled Bell pairs in which $\ket{\phi}$ contributes to the dominant part of the entanglement entropy of $A$ and $\ket{\sigma}$ captures the quantum fluctuations,
\begin{equation}
\begin{aligned}
\ket{\phi}=&\sum_{m=0}^{e^{S(A)-O(\sqrt{S(A)})}}\ket{m\bar m}_{\alpha\bar{\alpha}},\\
\ket{\sigma}=&\sum_{n=0}^{e^{O(\sqrt{S(A)})}}\sqrt{\tilde{p}_{n\Delta}}\ket{n\bar n}_{\beta\bar{\beta}}
\end{aligned}
\end{equation}
with Hilbert space dimensions:
\begin{equation}
\begin{aligned}
\dim\mathcal{H}_{\alpha}=&e^{S(A)-O(\sqrt{S(A)})}\equiv \Delta,\\
\dim\mathcal{H}_{\beta}=&e^{O(\sqrt{S(A)})},\\
\dim(\mathcal{H}_{\alpha}\otimes\mathcal{H}_{\beta})=&e^{S(A)},
\end{aligned}
\end{equation}
where $\Delta$ is the size of a block in eigenvalue space, $\tilde{p}_{n\Delta}$ is the average eigenvalue of the block and $S(A)$ is the entanglement entropy of subsystem $A$.

\begin{figure}[htbp]
	\begin{center}
		\begin{minipage}{7cm}
			\includegraphics[height=7.5cm,clip]{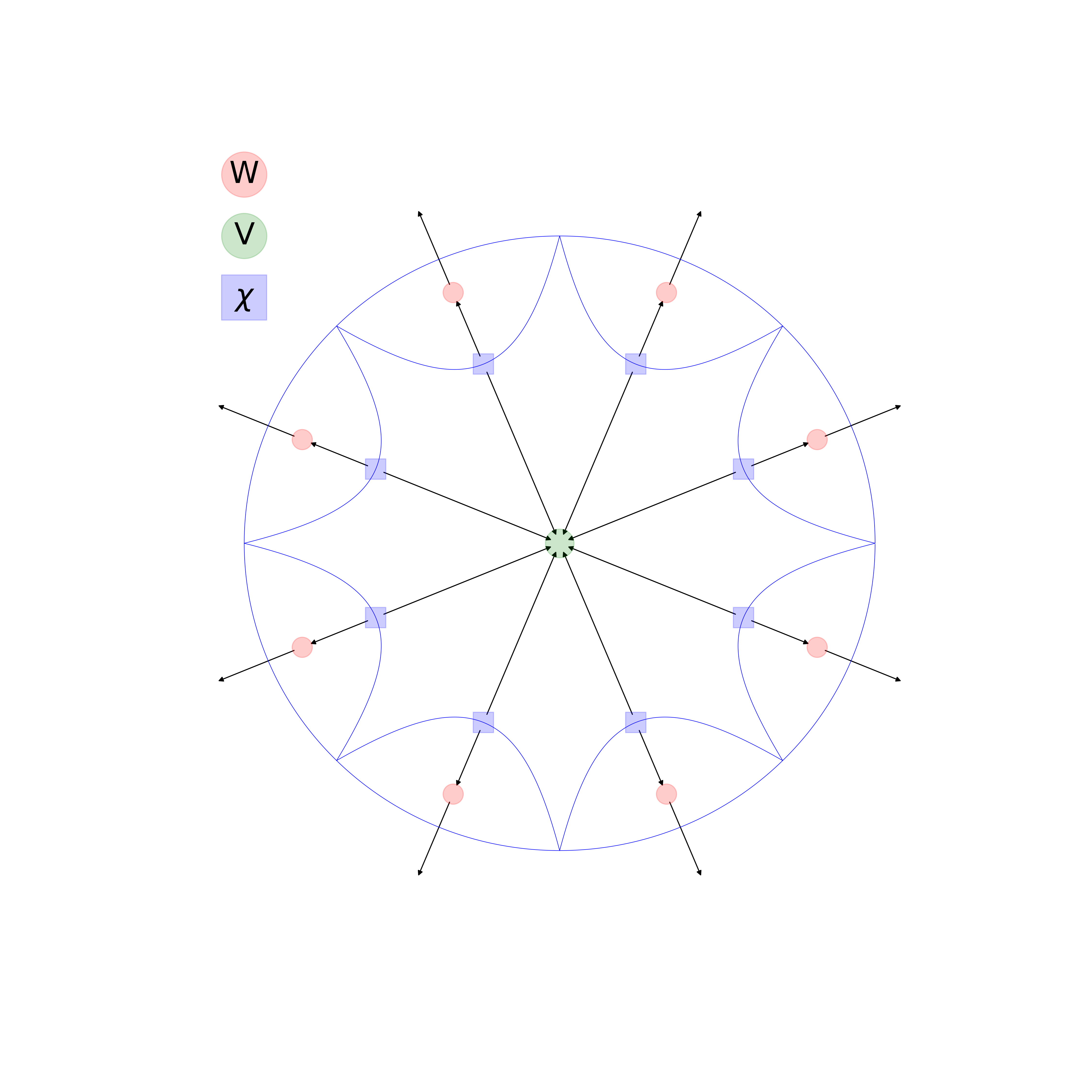}\\
			(a)
		\end{minipage}
		\begin{minipage}{7cm}
			\includegraphics[height=8cm,clip]{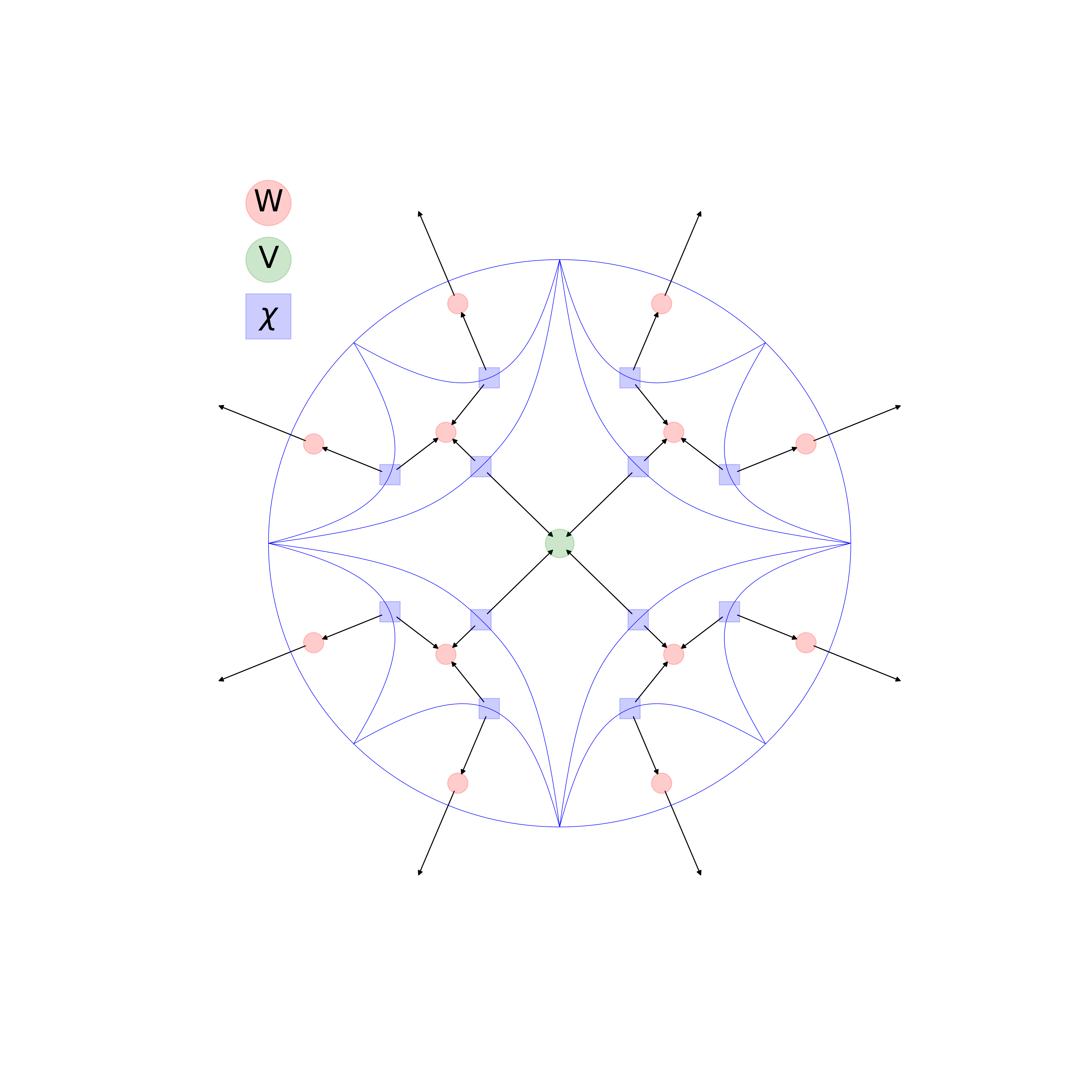}\\
			(b)
		\end{minipage}
		\caption{The first layer (a) and the second layer (b) of original surface growth scheme with $N=8$, and the surfaces $\chi$(the blue squares) always anchor on the boundary in this scheme.}
		\label{fig_original_8}
	\end{center}	
\end{figure}

Consider dividing the boundary into $N$ identical segments and using the RT surfaces of each segment to construct the first layer, the next layer is constructed in the similar way by taking the adjacent two segments into a new one. By iterating this process, a hierarchical structure of bulk minimal surfaces emerges. Explicitly, the first layer (Fig.\ref{fig_original_8}-a) of the bulk minimal surfaces (corresponding to $N$ segments) is described by the wave function
\begin{equation}
\ket{\Psi}^{(1)}=V_N'^{(1)}\prod_{i=1}^{N}W_i^{(1)}\ket{\phi}_i^{(1)}
\ket{\sigma}_i^{(1)},
\end{equation}
the second layer which contains $N/2$ RT surfaces (Fig.\ref{fig_original_8}-b) is described by boundary state
\begin{equation}
\ket{\Psi}^{(2)}=V_M'^{(2)}\left(\prod_{j=1}^{M=N/2}W_j'^{(2)}\ket{\phi}_j^{(2)}\ket{\sigma}_j^{(2)}\right)\left(\prod_{i=1}^{N}W_i^{(1)}
\ket{\phi}_i^{(1)}\ket{\sigma}_i^{(1)}\right),
\end{equation}
and subsequent layers refine the bulk geometry through progressive subdivisions, for example, the $k$-th layer is generated by $N/2^{k-1}$ segments
\begin{equation}
\ket{\Psi}^{(k)}=V'^{(k)}\prod_{j=1}^{k}\prod_{i=1}^{N/2^{k-1}}W_i'^{(j)}\ket{\phi}_i^{(j)}\ket{\sigma}_i^{(j)}.\label{kth}
\end{equation}
Each layer $k$ corresponds to a deeper radial slice of the bulk geometry, reflecting the multi-scale entanglement renormalization ansatz (MERA)~\cite{Bao:2018,Lin:2020ufd,miyaji:1503,Vidal:2008,vidal:1812,vidal:2007,hayden0204}. However, the OSED tensor network formalism imposes a constraint in eq.(\ref{kth}): the iterative subdivision process forces $N$ to reduce by half at each layer, thereby restricting the range of $k$. This restriction can be removed by extending the OSED tensor network which allows it to describe the general surface growth scheme.

\section{Tensor network description of the general surface growth scheme}
\label{sec-gen-sg}
In this section, we will extend the OSED tensor network to describe the general surface growth scheme in which the next layer of bulk minimal surface is anchored on the previous layer. In addition, the original tensor network construction of surface growth scheme only considered the pure state case, in which the bulk is the pure AdS spacetime, we will also generalize the framework by changing starting points of the surface growth and extend the OSED tensor network construction to include asymptotically AdS spacetime with black hole horizon.

\subsection{The generalized OSED tensor network}
The initial state of the boundary-horizon system is expressed as
\begin{equation}
\Psi^{(0)a^{(0)} b}=V_{AB}^{(0)}\Omega^{Bb}\chi_1^{(0)A_1 a_1^{(0)}}\cdots\chi_N^{(0)A_N a_N^{(0)}},
\end{equation}
where the indices $A$ and $a^{(0)}$ equivalently represent the series of indices $A_1\cdots A_N$ and $a^{(0)}_1\cdots a^{(0)}_N$, $a^{(0)}$ labels boundary segments, and $b$ labels the horizon. The tensor $\chi^{(k)}$ combines the RT surface states $\phi^{(k)}$ and $\sigma^{(k)}$ but $\chi^{(0)}$ is the initial boundary state, regarded as the combination of the RT surface states $\phi^{(0)}$ and $\sigma^{(0)}$ for simplicity. The surface growth operation is then applied by extending bulk minimal surfaces from the turning points of previous layers. For the first layer, this yields:
\begin{equation}
\begin{aligned}
\Psi^{(0)a^{(0)} b}=&V_{AB}^{(0)}\Omega^{Bb}\chi_1^{(0)A_1 a_1^{(0)}}\cdots\chi_N^{(0)A_N a_N^{(0)}}\\
=&V_{AB}^{(1)}\Omega^{Bb}\chi_1^{(1)A_1 a_1^{(1)}}\cdots\chi_N^{(1)A_N a_N^{(1)}}\\
&W_{a_1^{(1)} C_1^{(0)} D_2^{(0)}}^{(1)}W_{a_2^{(1)} C_2^{(0)} D_3^{(0)}}^{(1)}\cdots W_{a_N^{(1)} C_N^{(0)} D_1^{(0)}}^{(1)}\chi_1^{(0)C_1^{(0)} D_1^{(0)} a_1^{(0)}}\cdots\chi_N^{(0)C_N^{(0)} D_N^{(0)} a_N^{(0)}}\\
\equiv&V_{AB}^{(1)}\Omega^{Bb}\chi_1^{(1)A_1 a_1^{(1)}}\cdots\chi_N^{(1)A_N a_N^{(1)}}\tilde{W}_{a^{(1)}}^{(1)(0)a^{(0)}},
\end{aligned}\label{1stlayer}
\end{equation}
where $\tilde{W}^{(1)(0)}$ encodes the isometric transition between the first layer and the AdS boundary. The Hilbert space dimensions at the first layer are given by:
\begin{equation}
\begin{aligned}
\dim\mathcal{H}_{a^1}=&e^{NS^{(1)}},\\
\dim\mathcal{H}_{b}=&e^{S_h},
\end{aligned}
\end{equation}
where $S^{(1)}$ is the entanglement entropy of a RT surface in the first layer, and $S_h$ is the entanglement entropy of the horizon. With eq.(\ref{1stlayer}) the first-layer state of the surface-horizon system can be expressed as
\begin{equation}
\Psi^{(1)a^{(1)}b}\equiv V_{AB}^{(1)}\Omega^{Bb}\chi_1^{(1)A_1 a_1^{(1)}}\cdots\chi_N^{(1)A_N a_N^{(1)}},
\end{equation}
such that eq.(\ref{1stlayer}) can be rewritten as
\begin{equation}
\Psi^{(0)a^{(0)} b}=\Psi^{(1)a^{(1)}b}\tilde{W}_{a^{(1)}}^{(1)(0)a^{(0)}}.
\end{equation}

Next, repeating this process, the tensor network of the second layer state can be expressed as:
\begin{equation}
\begin{aligned}
\Psi^{(0)a^0 b}=&V_{AB}^{(1)}\Omega^{Bb}\chi_1^{(1)A_1 a_1^{(1)}}\cdots\chi_N^{(1)A_N a_N^{(1)}}\\
&W_{a_1^{(1)} C_1^{(0)} D_2^{(0)}}^{(1)}W_{a_2^{(1)} C_2^{(0)} D_3^{(0)}}^{(1)}\cdots W_{a_N^{(1)} C_N^{(0)} D_1^{(0)}}^{(1)}\chi_1^{(0)C_1^{(0)} D_1^{(0)} a_1^{(0)}}\cdots\chi_N^{(0)C_N^{(0)} D_N^{(0)} a_N^{(0)}}\\
=&V_{AB}^{(2)}\Omega^{Bb}\chi_1^{(2)A_1 a_1^{(2)}}\cdots\chi_N^{(2)A_N a_N^{(2)}}\\
&W_{a_1^{(2)} C_1^{(1)} D_2^{(1)}}^{(2)}W_{a_2^{(2)} C_2^{(1)} D_3^{(1)}}^{(2)}\cdots W_{a_N^{(2)} C_N^{(1)} D_1^{(1)}}^{(2)}\chi_1^{(1)C_1^{(1)} D_1^{(1)} a_1^{(1)}}\cdots\chi_N^{(1)C_N^{(1)} D_N^{(1)} a_N^{(1)}}\\
&W_{a_1^{(1)} C_1^{(0)} D_2^{(0)}}^{(1)}W_{a_2^{(1)} C_2^{(0)} D_3^{(0)}}^{(1)}\cdots W_{a_N^{(1)} C_N^{(0)} D_1^{(0)}}^{(1)}\chi_1^{(0)C_1^{(0)} D_1^{(0)} a_1^{(0)}}\cdots\chi_N^{(0)C_N^{(0)} D_N^{(0)} a_N^{(0)}}\\
\equiv&\Psi^{(2)a^{(2)} b}\tilde{W}_{a^{(2)}}^{(2)(1)a^{(1)}}\tilde{W}_{a^{(1)}}^{(1)(0)a^{(0)}},
\end{aligned}
\end{equation}
where $\tilde{W}^{(2)(1)}$ encodes the isometric transition between the second layer and the first layer, and $\Psi^{(2)a^{(2)} b}$ is the second layer state of the surface-horizon system:
\begin{equation}
\Psi^{(2)a^{(2)}b}=V_{AB}^{(2)}\Omega^{Bb}\chi_1^{(2)A_1 a_1^{(2)}}\cdots\chi_N^{(2)A_N a_N^{(2)}}.
\end{equation}
The Hilbert space dimensions at the second layer are given by
\begin{equation}
\dim\mathcal{H}_{a^{(2)}}=e^{NS^{(2)}}.
\end{equation}

\begin{figure}[htbp]
	\begin{center}
	\begin{minipage}{9cm}
		\includegraphics[width=7cm,clip]{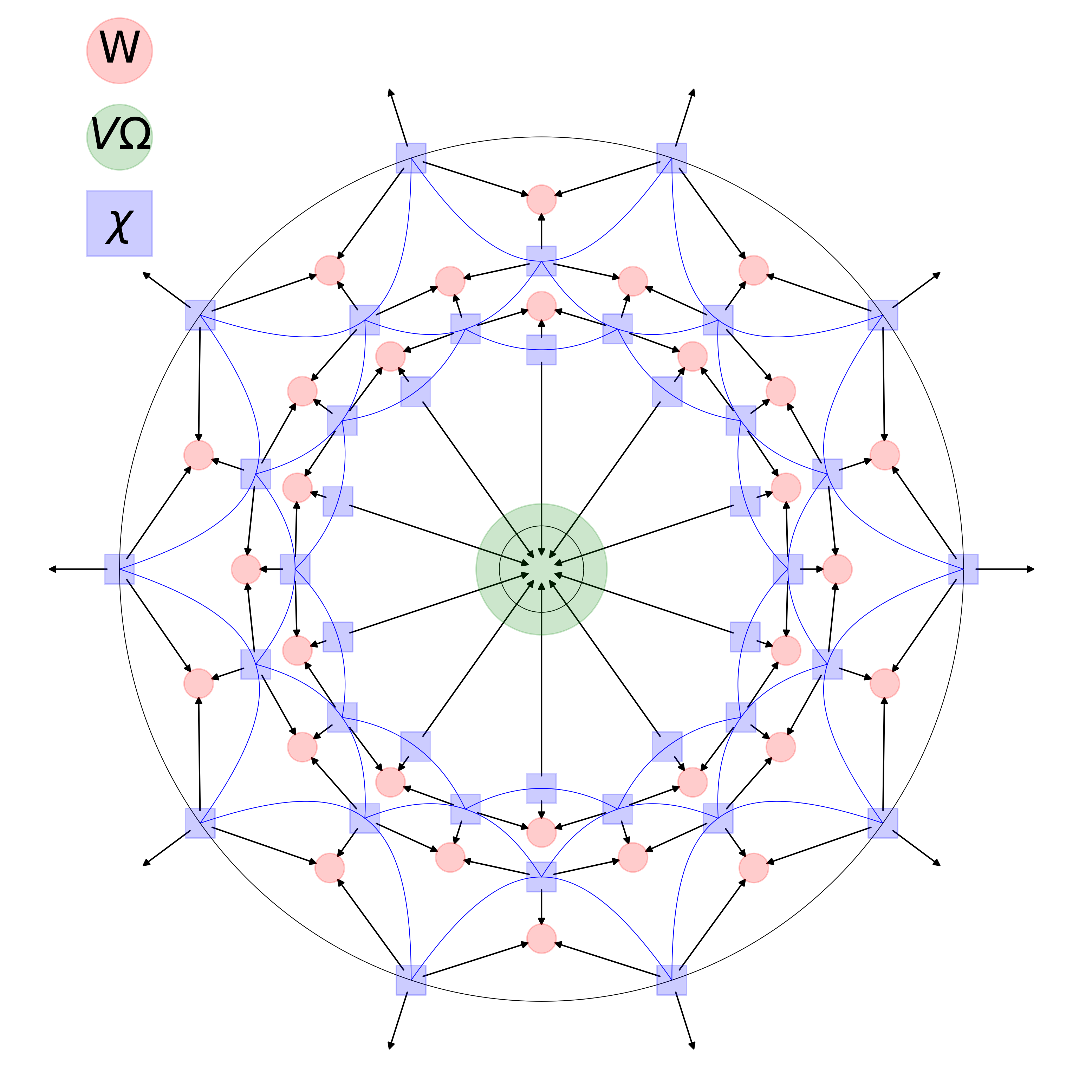}
	\end{minipage}
	\begin{minipage}{4cm}
		\includegraphics[width=3.5cm,clip]{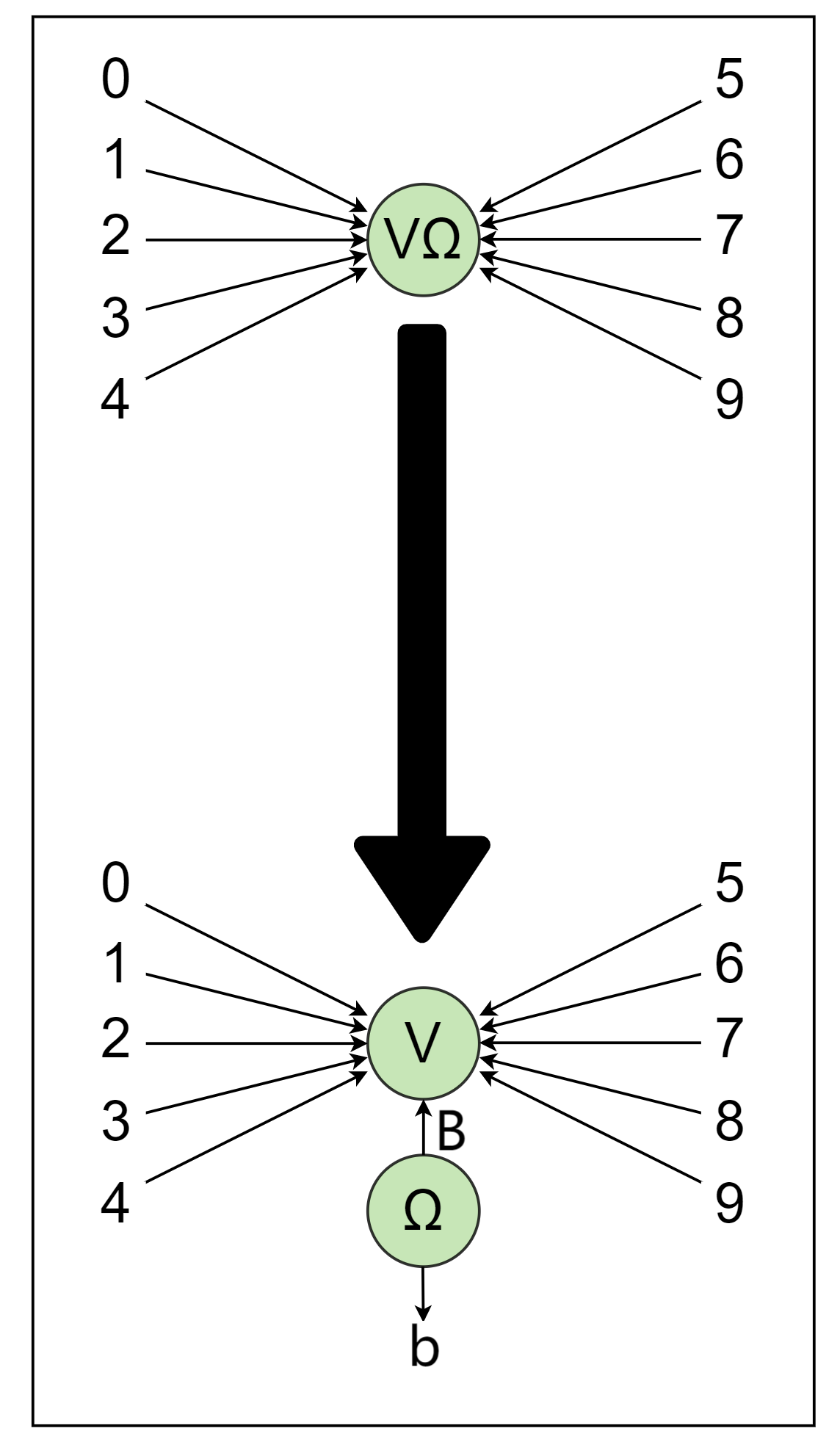}
	\end{minipage}
	\caption{The left is the tensor network of general surface growth scheme with $N=10$. The $V\Omega$ (the green circle) includes the bulk spacetime $V$ and the horizon $\Omega$, which have hidden labels and are shown in the right image. In this scheme, the surfaces growing out anchor on the turning points of previous surfaces.}
	\label{fig_general_5}
	\end{center}	
\end{figure}

Similarly, repeating this process generates higher layer states. For the $k$-th layer:
\begin{equation}
\begin{aligned}
\Psi^{(0)a^{(0)} b}=&\Psi^{(k)a^{(k)} b}\tilde{W}_{a^{(k)}}^{(k)(k-1)a^{(k-1)}}\cdots\tilde{W}_{a^{(1)}}^{(1)(0)a^{(0)}}.
\end{aligned}
\end{equation}
where $\tilde{W}$ encodes the isometric transition between layers, and $\Psi^{(k)a^{(k)} b}$ is the $k$-th layer state of the surface-horizon system:
\begin{equation}
\Psi^{(k)a^{(k)} b}=V_{AB}^{(k)}\Omega^{Bb}\chi_1^{(k)A_1 a_1^{(k)}}\cdots\chi_N^{(k)A_N a_N^{(k)}}.
\end{equation}
The Hilbert space dimensions at the $k$-th layer are given by
\begin{equation}
\dim\mathcal{H}_{a^{(k)}}=e^{NS^{(k)}}.
\end{equation}

Crucially, the subdivision count $N$ remains fixed across layers, eliminating the original constraint. This allows $k$ to take any positive integer value, enabling lasting growth of the surfaces toward the horizon without dependence on $N$ (Fig.\ref{fig_general_5}).

\subsection{The limit of the surface growth}
Geometrically, the iterative surface growth process drives the constructed minimal surfaces asymptotically toward the center of the bulk spacetime. It has been shown in~\cite{Yu:2020zwk} that  these surfaces will approach to the horizon after finite steps of growth. In this section, we rigorously demonstrate this behavior through a monotonicity argument and entropy constraints, and show that the surface growth will reach the black hole horizon in the $k\rightarrow\infty$ limit.

Consider the $i$-th minimal surface in the $k$-th layer. The Hilbert space dimension associated with the index $a^{(k)}_i$ is given by:
\begin{equation}
\dim\mathcal{H}_{a^{(k)}_i}=e^{S^{(k)}},
\end{equation}
where $S^{(k)}$ denotes the entanglement entropy of the $k$-th layer surface. Correspondingly, the Hilbert space dimension of the composite indices $C^{(k-1)}_i D^{(k-1)}_{i+1}$ from the previous layer is
\begin{equation}
\dim\mathcal{H}_{C^{(k-1)}_i D^{(k-1)}_{i+1}}=e^{\frac{1}{2}S^{(k-1)}}e^{\frac{1}{2}S^{(k-1)}}=e^{S^{(k-1)}}.
\end{equation}

Because of the minimal surface growing, the isometric tensor $W_{a_i^{(k)} C_i^{(k-1)} D_{i+1}^{(k-1)}}^{(k)}$ reduces the dimension to $e^{S^(k)}$, implying the entropy inequality:
\begin{equation}
S^{(k)}< S^{(k-1)},\ \ \forall k\in\mathbb{N}_+.
\end{equation}

Setting $N=1$ when constructing the $(k+1)$-th layer surfaces, the new surface will coincide with the horizon and the tensor network will be expressed as
\begin{equation}
\begin{aligned}
\Psi^{(k)a^{(k)} b}=&V_{AB}^{(k)}\Omega^{Bb}\chi_1^{(k)A_1 a_1^{(k)}}\cdots\chi_N^{(k)A_N a_N^{(k)}}\\
=&V_{AB}^{(h)}\Omega^{Bb}\chi^{(h)A a^{(h)}}\\
&W^{(h)}_{a^{(h)} C^{(k)}}\chi_1^{(k)C_1^{(k)} a_1^{(k)}}\cdots\chi_N^{(k)C_N^{(k)} a_N^{(k)}}.
\end{aligned}
\end{equation}
The dimensions of Hilbert space at the $(k+1)$-th layer are given by
\begin{equation}
\begin{aligned}
\dim\mathcal{H}_{a^{(h)}}=&e^{S_h},\\
\dim\mathcal{H}_{C^{(k)}}=&e^{NS^{(k)}}
\end{aligned}
\end{equation}
with the entropy inequality:
\begin{equation}
\frac{S_h}{N}\le S^{(k)}\le S^{(1)}.
\end{equation}
This monotonic decrease in entropy establishes that the sequence $\{S^{(k)}\}$ is bounded and strictly decreasing. By the Monotone Convergence Theorem, $S^{(k)}$ must converge to a finite limit as $k\rightarrow\infty$.

To identify this limit, suppose the surfaces in the $(k-1)$-th-layer surfaces coincide with the horizon. The $k$-th-layer surfaces also coincide with the horizon and the entanglement entropy of each segment is satisfied
\begin{equation}
S^{(k)}=S^{(k-1)}=\frac{S_h}{N}.
\end{equation}
Thus, the limit of the sequence $\{S^{(k)}\}$  converged to is:
\begin{equation}
\lim\limits_{k\rightarrow\infty}S^{(k)}=\frac{S_h}{N}.
\end{equation}
Consequently, in the limit $k\rightarrow\infty$, the iteratively grown surfaces converge to the horizon, completing the geometric reconstruction of the bulk spacetime, and the surface growth process dynamically encodes the transition from boundary entanglement to horizon physics.

\begin{figure}[htbp]
	\begin{center}
		\begin{minipage}{8cm}
			\includegraphics[height=9cm,clip]{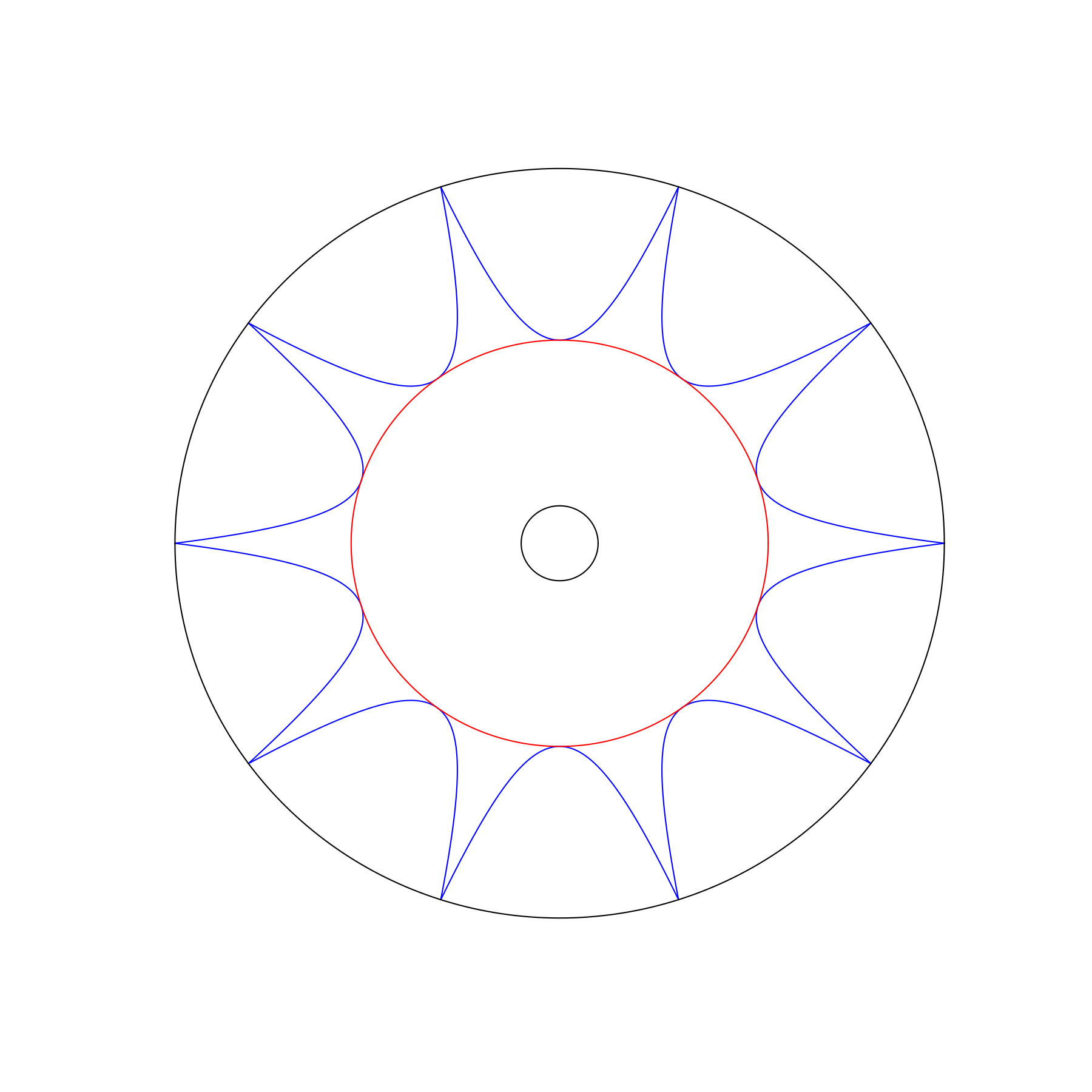}\\
			(a)
		\end{minipage}
		\begin{minipage}{8cm}
			\includegraphics[height=9cm,clip]{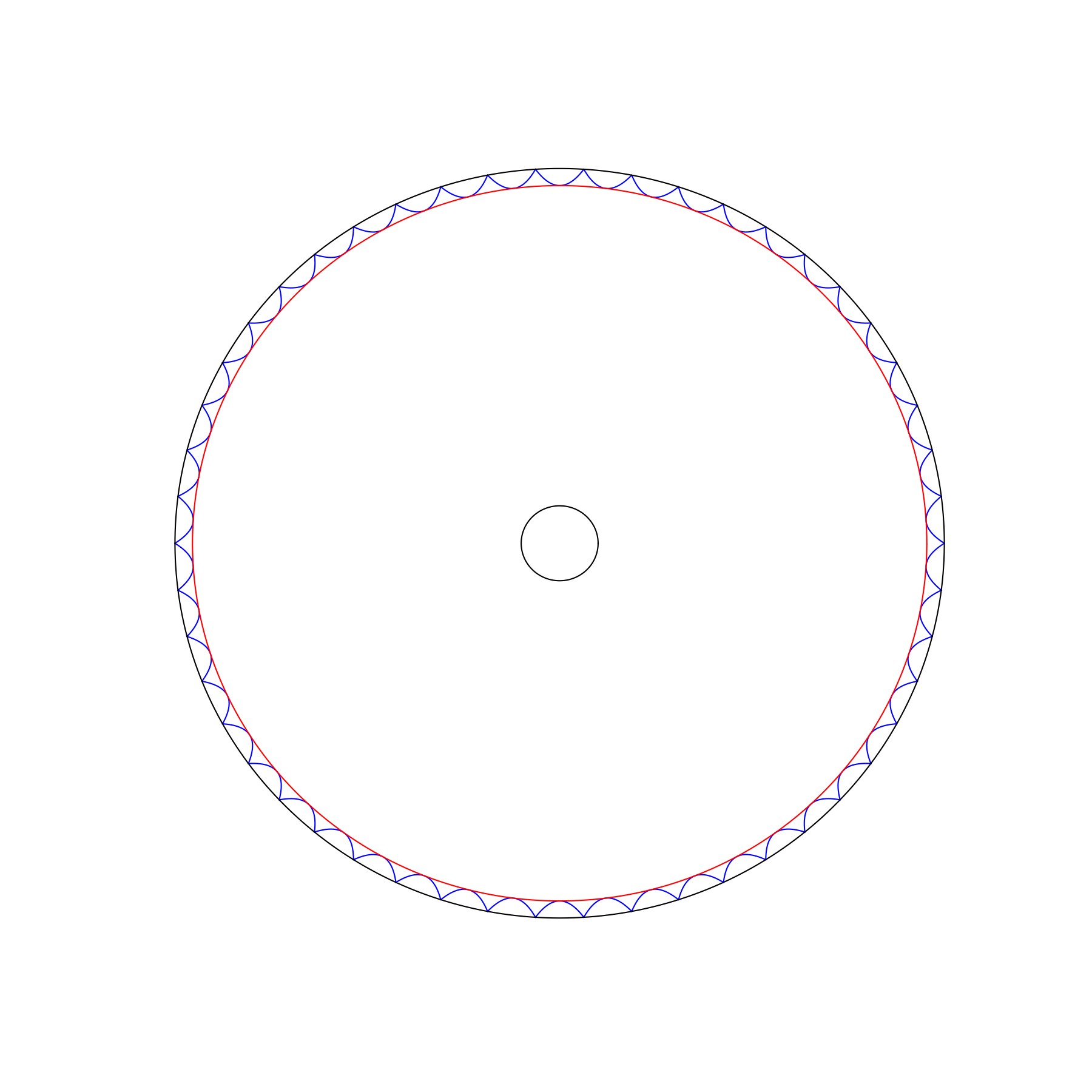}\\
			(b)
		\end{minipage}
		\caption{The partition number $N$ in (a) is chosen as $N=10$ and $N$ in (b) is chosen as $N=50$. In the continuum limit $N\rightarrow\infty$, a layer of homogenous and isotropic discrete bulk minimal surfaces converge to a smooth radial cutoff surface.}
		\label{fig_10_50}
	\end{center}
\end{figure}

To establish a smooth geometric correspondence, we analyze the continuum limit of the homogenous and isotropic surface growth as the partition number $N\rightarrow\infty$, as illustrated in Fig.\ref{fig_10_50}. This limit bridges the discrete tensor network construction with the continuous radial evolution of bulk geometry.

\subsection{Geometric convergence to radial surfaces}
Consider a non-rotating BTZ spacetime with metric:
\begin{equation}
ds^2=-\frac{r^2-M}{L^2}dt^2+\frac{L^2}{r^2-M}dr^2+r^2d\phi^2,
\end{equation}
whose horizon is located at $r_h=\sqrt{M}$. The surface growth scheme generates minimal surfaces governed by the equation~\cite{Yu:2020zwk}:
\begin{equation}
\Delta \phi = \phi-\phi_* = -\frac{L}{r_h}\ln\left(\sqrt{1-\frac{r_h^2}{r^2}}-\sqrt{\frac{r_h^2}{r_*^2}-\frac{r_h^2}{r^2}}\right)+\frac{L}{r_h}\ln\sqrt{1-\frac{r_h^2}{r_*^2}}\label{change_phi}
\end{equation}
with $(r_*,\phi_*)$ denoting the turning point of the surface.

Firstly, the first step of surface growth constructs a cut-off surface from $r\rightarrow\infty$ whose turning points are $r_*=r_0\gg r_h$, while the change of angle $\Delta\phi$ is
\begin{equation}
\Delta\phi=\frac{L}{2r_h}\ln\left(1+\frac{2r_h}{r_0-r_h}\right)=\frac{L}{r_0}+\mathcal{O}(r_0^{-3}).
\end{equation}
As long as $r_0\gg L$, the constructed surface has an infinitesimal angle range and the number of segments $N$ approaches infinity.

Secondly, we set $\Delta\phi = \frac{L}{r_0} \ll 1$, and one of the following steps induces a radial displacement from $r$ to $r_*$. With eq.(\ref{change_phi}), we obtain the change of radius
\begin{equation}
\Delta r=r_*-r=\frac{r(r_h^2-r^2)}{2r_0^2}+\mathcal{O}(r_0^{-3}).\label{change_r}
\end{equation}
In this step, the new surface has a non-radial deviation
\begin{equation}
\left|\frac{\Delta r}{r}\right|\simeq\frac{(r^2-r_h^2)}{2r_0^2}<\frac{1}{2},
\end{equation}
and when the step number is large enough i.e. $r\ll r_0$, the deviation
\begin{equation}
\left|\frac{\Delta r}{r}\right|\simeq\frac{(r^2-r_h^2)}{2r_0^2}\ll 1.
\end{equation}
Therefore, the layer of surfaces can be treated as a radial surface when $N\rightarrow\infty$.

\begin{figure}[htbp]
	\begin{center}
		\includegraphics[height=9cm,clip]{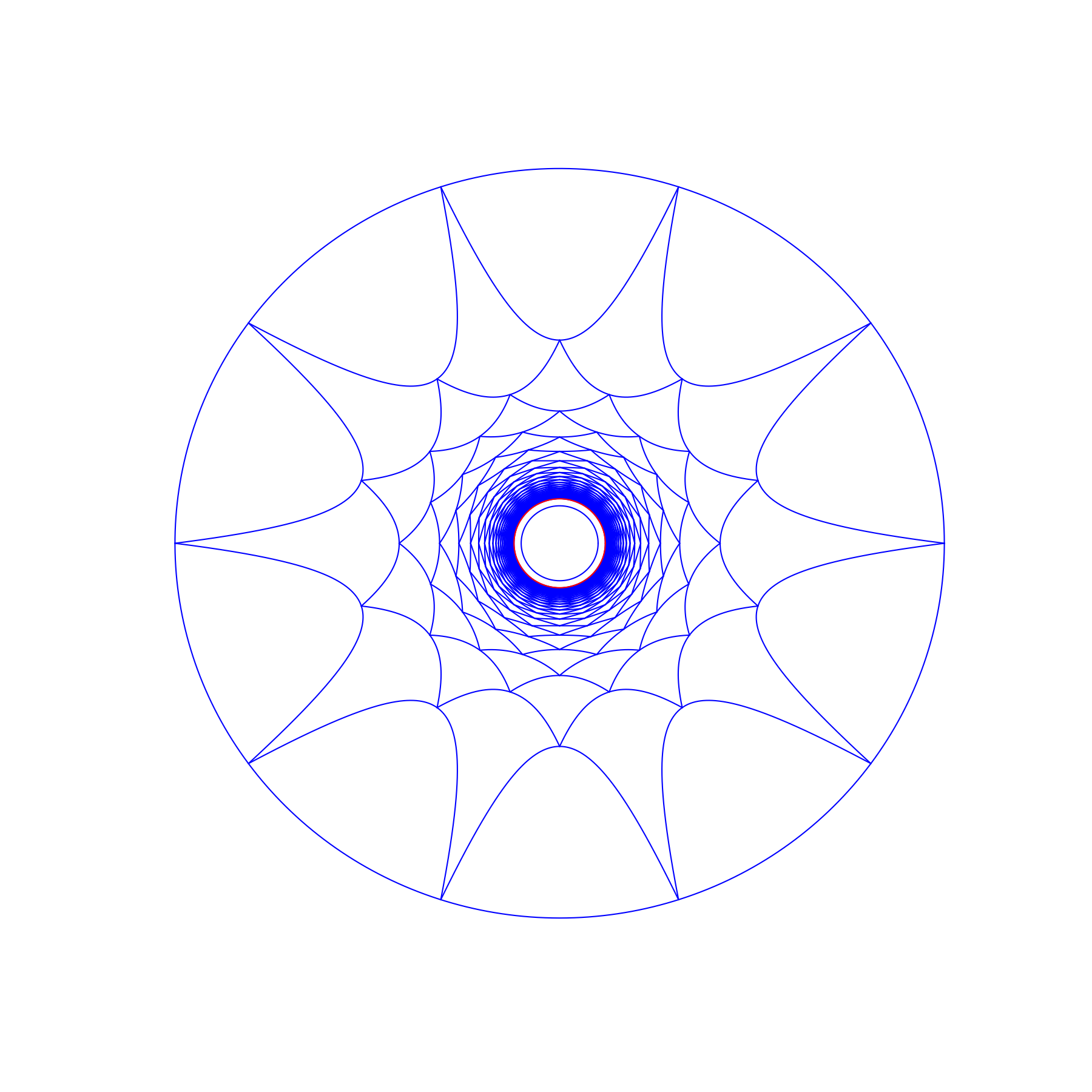}
		\caption{Homogenous and isotropic surface growth from the cutoff surface $r_0$ to the horizon, and the growing step number shown in the image is 50.}
		\label{fig_final}
	\end{center}
\end{figure}

Finally, to confirm the range of surface growth, we change the eq.(\ref{change_r}) to a differential equation with an initial condition
\begin{equation}
\Delta r\sim\frac{dr}{dx}=\frac{r(r_h^2-r^2)}{2r_0^2},\ \ r|_{x=0}=r_0,
\end{equation}
where $dx$ is one step of surface growth. And the solution is
\begin{equation}
r=\frac{e^{\frac{r_h^2 x}{2r_0^2}}r_h r_0}{\sqrt{e^{\frac{r_h^2 x}{r_0^2}}r_0^2-r_0^2+r_h^2}}.
\end{equation}
Subsequently, taking the limit $x\rightarrow +\infty$, we find:
\begin{equation}
r|_{x\rightarrow +\infty}=r_h.
\end{equation}
This implies that the surface growth scheme dynamically fills the bulk spacetime from the cutoff surface $r_0$ to the horizon $r_h$, covering the radial domain $(r_0,+\infty)$ in the continuum limit $N\rightarrow\infty$. An example of $N=10$ with $50$ growing steps is shown in Fig.\ref{fig_final}.

\subsection{Continuum state evolution}
The $k$-th layer state is expressed as:
\begin{equation}
\Psi^{(0)a^{(0)} b}=\Psi^{(k)a^{(k)} b}\tilde{W}_{a^{(k)}}^{(k)(k-1)a^{(k-1)}}\cdots\tilde{W}_{a^{(1)}}^{(1)(0)a^{(0)}},
\end{equation}
or equivalently,
\begin{equation}
\ket{\Phi_V}^{(0)}=\tilde{W}^{(k)(k-1)}\cdots\tilde{W}^{(1)(0)}\ket{\Phi_V}^{(k)}.
\end{equation}
In the limit $N\rightarrow\infty$, the tensors $\tilde{W}$ and the step number $k$ are replaced by an evolution operator $U$ and the radial parameter $\mu$:
\begin{equation}
\ket{\Phi(\mu)}=U(\mu,0)\ket{\Phi(0)}
\end{equation}
and the evolution operator $U$ satisfies rules:
\begin{equation}
U(a,b)U(b,c)=U(a,c)\ ,\ U(a,b)=U^{\dagger}(b,a)\ ,\ U(a,b)U(b,a)=1.
\end{equation}
The state of the surface-horizon system $\ket{\Phi(\mu)}$ decomposes into a radial surface $\ket{V(\mu)}$ and the horizon $\ket{\Omega}$:
\begin{equation}
\ket{\Phi(\mu)}=M(\mu)\ket{V(\mu)}\ket{\Omega},
\end{equation}
where $M(\mu)$ encodes metric data. It is obvious that the evolution operator $U$ does not change the horizon state $\ket{\Omega}$,
\begin{equation}
\begin{aligned}
\ket{\Phi(\mu_1)}=&M(\mu_1)\ket{V(\mu_1)}\ket{\Omega}\\
=&U(\mu_1,\mu_2)\ket{\Phi(\mu_2)}\\
=&U(\mu_1,\mu_2)M(\mu_2)\ket{V(\mu_2)}\ket{\Omega}
\end{aligned}
\end{equation}
and the density operators have the same properties
\begin{equation}
\begin{aligned}
\rho_{V}(\mu_1)=&U(\mu_1,\mu_2)\rho_{V}(\mu_2)U^{\dagger}(\mu_1,\mu_2),\\
\rho_{\Omega}=&U(\mu_1,\mu_2)\rho_{\Omega}U^{\dagger}(\mu_1,\mu_2),
\end{aligned}
\end{equation}
where $\rho_{V}(\mu)$ is the density operator at the surfaces arriving at parameter $\mu$, and $\rho_{\Omega}$ is the density operator at the horizon. Considering a operator $\mathcal{O}$ at growing surfaces, its expectation value is
\begin{equation}
\text{Tr}\left(\rho_{V}(\mu)\mathcal{O}\right)=\text{Tr}\left(\rho_{V}(0)U^{\dagger}(\mu,0)\mathcal{O}U(\mu,0)\right).\label{evO}
\end{equation}

Specially, when the surfaces arrives at the horizon i.e. $\mu=\mu_h$, the $M$ is a diagonal matrix
\begin{equation}
\begin{aligned}
\ket{\Phi(\mu_h)}=&M(\mu_h)\ket{V(\mu_h)}\ket{\Omega}=\sum_{i}\lambda_i\ket{V(\mu_h)}_i\ket{\Omega}_i,
\end{aligned}
\end{equation}
for $S(\rho_{V})=S(\rho_{\Omega})=S_h$.

\section{Establish the connection between surface growth and $T\bar{T}$ deformation}
\label{sec-TTbar}
In this section, we will establish the connection between the surface growth scheme and $T\bar{T}$ deformation of CFT$_2$. Let us first recall the important properties of $T\bar{T}$ deformation. From the RG perspective, the $ T\bar{T}$ operator is irrelevant, inducing significant modifications to the UV dynamics of the deformed CFT. Holographically, such deformations are conjectured to alter the IR geometry of the bulk spacetime. Notably, for negative coupling constants, the $T\bar{T}$-deformed CFT has been suggested to be dual to gravitational physics in AdS$_3$ with a finite radial cutoff~\cite{McGough:2016lol}.

An alternative formulation suggested that the $T\bar{T}$-deformed CFT admits a holographic dual description as AdS$_3$ governed by mixed boundary conditions~\cite{Guica:2019nzm}. Crucially, in pure Einstein gravity, the finite-cutoff and mixed-boundary-condition dualities exhibit mathematical equivalence rigorously proven through bulk diffeomorphism invariance, thereby establishing their physical consistency within the context of AdS$_3$/CFT$_2$ correspondence. Therefore, the holographic duality with $T\bar{T}$ deformation modifies the conventional framework of AdS/CFT correspondence, which prompts a critical inquiry into the mechanism of spacetime emergence within this generalized correspondence. By using the surface/state correspondence~\cite{Miyaji:2015fia,Miyaji:2015yva}, this problem has been investigated in \cite{Chen:2019mis}, which proposed that quantum states of $T\bar{T}$ deformed CFT correspond to finite radial cutoff surface of the bulk AdS spacetime. On the other hand, note that the iterative growth of bulk minimal surfaces which describes the radial evolution in AdS spacetime, may directly maps to the operator-driven deformation flow governed by the $T\bar T$ perturbation. In other words, it is expected that the $T\bar T$ deformation of boundary CFT can provide a dynamical mechanism for specific configuration of the surface growth.

\subsection{$T\bar{T}$ deformation and the surface growth in asymptotically AdS$_3$ spacetime}
In what follows, we will establish the connection between the surface growth scheme and the $T\bar{T}$ deformation by adopting the mixed boundary conditions formalism. Considering a spacetime with a negative cosmological constant, the solution of Einstein's equations can be expressed by the Fefferman-Graham (FG) expansion~\cite{Henningson:1998gx,Balasubramanian:1999re}
\begin{equation}
\begin{aligned}
ds^2=&l^2\frac{d\rho^2}{4\rho^2}+g_{ab}(\rho,x^a)dx^adx^b,\\
g_{ab}(\rho,x^a)=&\frac{1}{\rho}\left(g_{ab}^{(0)}(x^a)+\rho g_{ab}^{(2)}(x^a)+\rho^2 g_{ab}^{(4)}(x^a)+\cdots\right),
\end{aligned}
\end{equation}
where $l$ is the curvature radius of the AdS spacetime. Specially, the expansion truncates at the second order in three dimensions. In this case, we have
\begin{equation}
g_{ab}^{(2)}=8\pi Gl(T_{ab}-g_{ab}^{(0)}T_c^c)\equiv 8\pi Gl\hat{T}_{ab},
\end{equation}
where $T_{ab}$ is the Brown-York stress tensor on the AdS boundary and
\begin{equation}
 g^{(4)}_{ab}=\frac{1}{4}g^{(2)}_{ac}g^{(0)cd}g^{(2)}_{bd}.
\end{equation}

To match the radial surface, we define the $T\bar{T}$ deformation by adding the $T\bar T$ operator to the Einstein-Hilbert action
\begin{equation}
\frac{\partial}{\partial\mu}I_{\mu}=-\frac{1}{2}\int d^2 x\sqrt{\gamma}\mathcal{O}_{T\bar T},
\end{equation}
in which $\mathcal{O}_{T\bar T}=T^{ab}T_{ab}-(T_c^{c})^2$. \omits{The variation of the Euclidean QFT action with respect to $\gamma_{ab}$ is
\begin{equation}
\delta I_{\mu}=\frac{1}{2}\int d^2 x\sqrt{\gamma}T_{ab}\delta\gamma^{ab}.
\end{equation}
}
When $\mu$ is infinitesimally varied, the change in $\sqrt{\gamma}\mathcal{O}_{T\bar T}$ and $T_{ab}$ is given by the flow equation
\begin{equation}
\frac{\partial}{\partial\mu}(\delta I)=-\delta(\frac{\partial}{\partial\mu}I)\Leftrightarrow \frac{\partial}{\partial\mu}(\sqrt{\gamma}T_{ab}\delta\gamma^{ab})=\delta(\sqrt{\gamma}\mathcal{O}_{T\bar T}),\label{eq_flow}
\end{equation}
\omits{Introducing the quantity and then the solutions of eq.(\ref{eq_flow}) are
\begin{equation}
\frac{\partial}{\p$\hat{T}_{ab}\equiv T_{ab}-\gamma_{ab}T_c^c$artial\mu}\gamma_{ab}=-2\hat{T}_{ab}, \ \frac{\partial}{\partial\mu}\hat{T}_{ab}=-\hat{T}_{ac}\gamma^{cd}\hat{T}_{db},\ \frac{\partial}{\partial\mu}(\hat{T}_{ac}\gamma^{cd}\hat{T}_{db})=0
\end{equation}
}
which can be solved as~\cite{Guica:2019nzm}
\begin{equation}
\begin{aligned}
\gamma^{(\mu)}_{ab}=&\gamma^{(0)}_{ab}-2\mu\hat{T}_{ab}^{(0)}+\mu^2\hat{T}_{ac}^{(0)}\gamma^{cd}_{(0)}\hat{T}_{db}^{(0)},\\
\hat{T}_{ab}^{(\mu)}=&\hat{T}_{ab}^{(0)}-\mu\hat{T}_{ac}^{(0)}\gamma^{cd}_{(0)}\hat{T}_{db}^{(0)}.
\end{aligned}
\end{equation}
where $\hat{T}_{ab}\equiv T_{ab}-\gamma_{ab}T_c^c$. By comparing with FG expansion in asymptotically AdS$_3$ spacetime, the $T\bar{T}$ deformation of CFT$_2$ can be identified with the radial cutoff surface of AdS$_3$ as
\begin{equation}
\begin{aligned}
g^{(0)}_{ab}=&\gamma^{(0)}_{ab},\\
g^{(2)}_{ab}=&8\pi Gl\hat{T}_{ab}\equiv 2C\hat{T}_{ab},\\
g^{(4)}_{ab}=&C^2\hat{T}_{ac}^{(0)}\gamma^{cd}_{(0)}\hat{T}_{db}^{(0)},\\
\rho_c=&-\frac{\mu}{C}.
\end{aligned}
\end{equation}

\subsection{Connection of operators between $T\bar{T}$ deformation and surface growth}
To establish the operator connection between surface growth dynamics and $T\bar{T}$ deformation, we construct the evolution operator formalism governing hierarchical entanglement structure propagation. Within the surface growth scheme, the metric operator $\underline{\gamma_{ab}}$ and reduced stress tensor operator $\underline{\hat{T}_{ab}}$ are defined through the corresponding expectation values in $T\bar{T}$-deformed CFT$_2$:
\begin{equation}
\begin{aligned}
\text{Tr}\left(\rho_{V}(0)\underline{\hat{T}}_{ab}\right)=&\hat{T}^{(0)}_{ab}\ \ ,\ \ \text{Tr}\left(\rho_{V}(0)\underline\gamma_{ab}\right)=\gamma^{(0)}_{ab},\\
\text{Tr}\left(\rho_{V}(\mu)\underline\gamma_{ab}\right)=&\gamma^{(\mu)}_{ab}=\gamma^{(0)}_{ab}-2\mu\hat{T}^{(0)}_{ab}+\mu^2\hat{T}^{(0)}_{ac}\gamma_{(0)}^{cd}\hat{T}^{(0)}_{db},\\
\text{Tr}\left(\rho_{V}(\mu)\underline{\hat{T}}_{ab}\right)=&\hat{T}^{(\mu)}_{ab}=\hat{T}^{(0)}_{ab}-\mu\hat{T}^{(0)}_{ac}\gamma_{(0)}^{cd}\hat{T}^{(0)}_{db},
\end{aligned}
\end{equation}
where $\hat{T}_{ab}^{(\mu)}=\ev{\hat{T}_{ab}}^{(\mu)}$ is the expectation value of $\hat{T}_{ab}$ in $T\bar{T}$-deformed CFT$_2$, and $\gamma_{ab}^{(\mu)}$ is the deformed metric, while $\text{Tr}\left(\rho_{V}(\mu)\underline\gamma_{ab}\right)$ and $\text{Tr}\left(\rho_{V}(\mu)\underline{\hat{T}}_{ab}\right)$ are the expectation values defined by eq.(\ref{evO}) at the growing surfaces, and $\rho_{V}(\mu)$ is the density matrix operator of the surfaces that grow out.

Furthermore, this definition requires that
\begin{equation}
\text{Tr}\left(\rho_{V}(0)\underline{\hat{T}}_{ac}\underline{\gamma}^{cd}\underline{\hat{T}}_{db}\right)=\ev{\hat{T}_{ac}\gamma^{cd}\hat{T}_{db}}^{(0)}.
\end{equation}
With Zamolodchikov's factorization formula in CFT$_2$
\begin{equation}
\ev{\hat{T}_{ac}\gamma^{cd}\hat{T}_{db}}^{(0)}=\ev{\hat{T}_{ac}}^{(0)}\gamma^{cd}_{(0)}\ev{\hat{T}_{db}}^{(0)},
\end{equation}
we can obtain the same formula in surface growth scheme
\begin{equation}
\text{Tr}\left(\rho_{V}(0)\underline{\hat{T}}_{ac}\underline{\gamma}^{cd}\underline{\hat{T}}_{db}\right)=\text{Tr}\left(\rho_{V}(0)\underline{\hat{T}}_{ac}\right)\text{Tr}\left(\rho_{V}(0)\underline{\gamma}^{cd}\right)\text{Tr}\left(\rho_{V}(0)\underline{\hat{T}}_{db}\right).
\end{equation}
Similarly, according to $\text{Tr}\left(\rho_{V}(\mu)\underline\gamma_{ac}\underline\gamma^{cb}\right)=\delta_a^b$, we can also request that the metric operators are satisfied with $\underline{\gamma}_{ac}\underline{\gamma}^{cb}=\underline{\mathbf{1}}\delta_a^b$ and $\underline{\gamma}^{ac}\underline{\gamma}_{cb}=\underline{\mathbf{1}}\delta_b^a$.

The unitary evolution operator $U(\mu,0)$ implements radial flow through path ordered exponentiation:
\begin{equation}
U(a,b)=\mathcal{P}e^{-i\int_b^a A(x)dx}
\end{equation}
and
\begin{equation}
U^{\dagger}(a,b)=\mathcal{P}^*e^{i\int_b^a A(x)dx},
\end{equation}
where $A^{\dagger}(x)=A(x)$ is Hermitian, and the symbols $\mathcal{P}$ and $\mathcal{P}^*$ mean the reverse path-ordering. By moving to the "Heisenberg picture", the evolution of $\underline{\gamma}_{ab}$ and $\underline{\hat{T}}_{ab}$ is given by
\begin{equation}
\begin{aligned}
U^{\dagger}(\mu,0)\underline{\gamma}_{ab} U(\mu,0)=&\underline{\gamma}_{ab}-2\mu\underline{\hat{T}}_{ab}+\mu^2\underline{\hat{T}}_{ac}\underline{\gamma}^{cd}\underline{\hat{T}}_{db},\\
U^{\dagger}(\mu,0)\underline{\hat{T}}_{ab}U(\mu,0)=&\underline{\hat{T}}_{ab}-\mu\underline{\hat{T}}_{ac}\underline{\gamma}^{cd}\underline{\hat{T}}_{db}.
\end{aligned}\label{UOU=O}
\end{equation}
Considering the $\mu$ term of eq.(\ref{UOU=O}), The commutator structure $\left[\int_0^{\mu}A(x)dx,\underline{\gamma}_{ab}\right]$ and $\left[\int_0^{\mu}A(x)dx,\underline{\hat{T}}_{ab}\right]$ are given by (please see appendix A for more details)
\begin{equation}
\begin{aligned}
\left[\int_0^{\mu}A(x)dx,\underline{\gamma}_{ab}\right]\equiv&2i\mu\underline{\hat{T}}_{ab},\\
\left[\int_0^{\mu}A(x)dx,\underline{\hat{T}}_{ab}\right]\equiv&i\mu\underline{\hat{T}}_{ac}\underline{\gamma}^{cd}\underline{\hat{T}}_{db}.\label{eqcom_0}
\end{aligned}
\end{equation}
With these formulas, we obtain (please see appendix B for more details)
\begin{equation}
\begin{aligned}
\left[\int_0^{\mu}A(x)dx,\underline{\gamma}^{ab}\right]=&-2i\mu\underline{\gamma}^{ac}\underline{\hat{T}}_{cd}\underline{\gamma}^{db},\\
\left[\int_0^{\mu}A(x)dx,\underline{\hat{T}}_{ac}\underline{\gamma}^{cd}\underline{\hat{T}}_{db}\right]=&0.
\end{aligned}\label{eqcom_1}
\end{equation}

Then considering the $\mu^2$ term of eq.(\ref{UOU=O}), we obtain (see appendix A)
\begin{equation}
\begin{aligned}
\tilde{\mathcal{P}}\left(\int_0^{\mu}dx_1\int_0^{\mu}dx_2\left[A(x_1),\left[A(x_2),\underline{\gamma}_{ab}\right]\right]\right)\equiv&2i\mu\left[\int_0^{\mu}A(x)dx,\underline{\hat{T}}_{ab}\right]=-2\mu^2\underline{\hat{T}}_{ac}\underline{\gamma}^{cd}\underline{\hat{T}}_{db},\\
\tilde{\mathcal{P}}\left(\int_0^{\mu}dx_1\int_0^{\mu}dx_2\left[A(x_1),\left[A(x_2),\underline{\hat{T}}_{ab}\right]\right]\right)\equiv&i\mu\left[\int_0^{\mu}A(x)dx,\underline{\hat{T}}_{ac}\underline{\gamma}^{cd}\underline{\hat{T}}_{db}\right]=0,
\end{aligned}
\end{equation}
where the symbol $\tilde{\mathcal{P}}$ means the same path order of commutation relations as $\mathcal{P}^*$. Because of the fact that the path-order does not effect the result and $\mu$ is arbitrary in the range of $\left[0,\mu_h\right]$, the operator $A(x)$ can be separated into $K(x)$ and $Q$:
\begin{equation}
A(x)=K(x)+Q
\end{equation}
and eq.(\ref{eqcom_0}) can be rewritten as
\begin{equation}\label{basic gamma That}
\begin{aligned}
\left[K(x),\underline{\gamma}_{ab}\right]=0\ ,&\ \left[Q,\underline{\gamma}_{ab}\right]=2i\underline{\hat{T}}_{ab},\\
\left[K(x),\underline{\hat{T}}_{ab}\right]=0\ ,&\ \left[Q,\underline{\hat{T}}_{ab}\right]=i\underline{\hat{T}}_{ac}\underline{\gamma}^{cd}\underline{\hat{T}}_{db},
\end{aligned}
\end{equation}
where $K(x)$ is the disentangler and $Q$ is the coarse grainer reducing the dimension of the labels in surface growth process.

Similarly, eq.(\ref{eqcom_1}) can be rewritten as
\begin{equation}\label{jinjie gamma1 tgt}
\begin{aligned}
\left[K(x),\underline{\gamma}^{ab}\right]=0\ ,&\ \left[Q,\underline{\gamma}^{ab}\right]=-2i\underline{\gamma}^{ac}\underline{\hat{T}}_{cd}\underline{\gamma}^{db},\\
\left[K(x),\underline{\hat{T}}_{ac}\underline{\gamma}^{cd}\underline{\hat{T}}_{db}\right]=0\ ,&\ \left[Q,\underline{\hat{T}}_{ac}\underline{\gamma}^{cd}\underline{\hat{T}}_{db}\right]=0.
\end{aligned}
\end{equation}

\begin{figure}[htbp]
	\begin{tikzpicture}
	\node [draw,rectangle](a)at(0,0){AdS$_3$, radial surfaces};
	\node [draw,rectangle](b)at(-3,-1){$T\bar{T}$ deformation, deformed by $\frac{\partial}{\partial\mu}$};
	\node [draw,rectangle](c)at(3,-1){Surface growth, growing with $Q$};
	\draw[->](a)--(b);
	\draw[->](b)--(a);
	\draw[->](a)--(c);
	\draw[->](c)--(a);
	\draw[->](c)--(b);
	\draw[->](b)--(c);
	\end{tikzpicture}
	\caption{The connection between surface growth scheme and $T\bar{T}$ deformation.}
\end{figure}
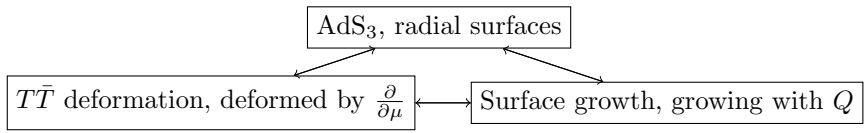

In conclusion, the $T\bar{T}$ deformation directly maps to the radial evolution step in the surface growth scheme. The commutator structure $\left[Q,\underline{\mathcal{O}}\right]$, where $Q$ generates the deformation of $\underline{\gamma}_{ab}$ and $\underline{\hat{T}}_{ab}$ encodes the bulk diffeomorphism invariance
\begin{equation}
i\left[Q,\underline{\mathcal{O}}\right]\sim\frac{\partial}{\partial\mu}\langle\mathcal{O}\rangle_{\mu}.
\end{equation}
Therefore, for the homogenous, isotropic and continuum case, the dynamical mechanism of iterative growth of bulk minimal surfaces in asymptotically AdS spacetime can be described by the radial RG flow driven by the $T\bar{T}$ deformation of boundary CFT, which provides a dynamical mechanism for the surface growth scheme and may shed light on reconstructing bulk gravitational dynamics via this scheme.

\section{Conclusions and discussions}\label{sect:conclusion}
In this paper, we extended the OSED tensor network to describe the more general surface growth scheme for bulk reconstruction in asymptotically AdS spacetime, which provided a state-to-geometry mapping between quantum states of tensor network and bulk minimal surfaces. We then analyzed the behavior of geometric convergence and continuum limit of the general surface growth scheme as the partition number $N\rightarrow\infty$, which bridges the tensor network construction with the continuous radial evolution of the bulk geometry. Finally, through an evolution operator formalism within the tensor network framework, we established an explicit connection between the surface growth process and $T\bar{T}$-deformed CFT$_2$. By implementing the operators $\underline{\gamma}_{ab}$ and $\underline{\hat{T}}_{ab}$ as fundamental building blocks in this scheme, we demonstrated systematic reconstruction of bulk geometry in asymptotically AdS$_3$ spacetime from the operators of the boundary CFT$_2$. Namely, the $T\bar{T}$ deformation of boundary CFT$_2$ can provide a dynamical mechanism for the surface growth in asymptotically AdS$_3$ spacetime.

Note that our present framework is only formulated within the AdS$_3$/CFT$_2$ correspondence, there are still many interesting problems to be studied such as the extension into higher dimensional spacetime. However, there will be many difficulties, for example, (i) resolution of non-trivial truncation in the FG expansion, and (ii) the non locality of stress tensors, since the absence of finite FG truncation renders metric and stress tensor deformations ill-defined under conventional $T\bar{T}$ operator flows in $d\ge 3$ dimensional spacetime. In addition, the current implementation of the operators $\underline{\gamma}_{ab}$ and $\underline{\hat{T}}_{ab}$ assumed homogenous and isotropic radial propagation, the extension into inhomogenous and anisotropic surface growth scheme and its connection with other operator deformation requires further investigation.

\section*{Acknowledgement}
We would like to thank S. He for useful discussions. J.R.S. was supported by the National Natural Science Foundation of China (No.~12475069) and Guangdong Basic and Applied Basic Research Foundation (No.~2025A1515011321). Y. S. was supported by the National Natural Science Foundation of China (No.12105113).

\begin{appendix}
\section*{APPENDIX}
\subsection*{APPENDIX A}
This appendix aims to give more details in deriving \eqref{eqcom_0}. Let us consider the evolution operator from $b$ to $a$ is defined as
\begin{equation}
\begin{aligned}
U(a,b)=&\mathcal{P}e^{-i\int_b^a A(x)dx}\\
=&1+(-i)\int_b^a dx_1 A(x_1)+\frac{(-i)^2}{2}\mathcal{P}\left(\int_b^a dx_1\int_b^a dx_2 A(x_1)A(x_2)\right)+\cdots\\
&+\frac{(-i)^n}{n!}\mathcal{P}\left(\int_b^a dx_1\cdots\int_b^a dx_n A(x_1)\cdots A(x_n)\right)+\cdots
\end{aligned}
\end{equation}
and we obtain
\begin{equation}
\begin{aligned}
U^{\dagger}(a,b)=&U(b,a)=\mathcal{P}e^{-i\int_a^b A(x)dx}\\
=&1+(-i)\int_a^b dx_1 A(x_1)+\frac{(-i)^2}{2}\mathcal{P}\left(\int_a^b dx_1\int_a^b dx_2 A(x_1)A(x_2)\right)+\cdots\\
&+\frac{(-i)^n}{n!}\mathcal{P}\left(\int_a^b dx_1\cdots\int_a^b dx_n A(x_1)\cdots A(x_n)\right)+\cdots\\
=&1+(i)\int_b^a dx_1 A(x_1)+\frac{(i)^2}{2}\mathcal{P}^*\left(\int_b^a dx_1\int_b^a dx_2 A(x_1)A(x_2)\right)+\cdots\\
&+\frac{(i)^n}{n!}\mathcal{P}^*\left(\int_b^a dx_1\cdots\int_b^a dx_n A(x_1)\cdots A(x_n)\right)+\cdots\\
=&\mathcal{P}^*e^{i\int_b^a A(x)dx},
\end{aligned}
\end{equation}
where $\mathcal{P}^*$ means the reverse path-ordering.

Considering an operator $\mathcal{O}$ changed with the evolution operator $U(a,b)$, we can write the terms of the expansion as
\begin{equation}
\begin{aligned}
U^{\dagger}(a,b)\mathcal{O}U(a,b):&\\
\mu^0\ \text{term}=&\mathcal{O},\\
\mu^1\ \text{term}=&(i)\int_b^a dx_1 A(x_1)\mathcal{O}+(-i)\mathcal{O}\int_b^a dx_1 A(x_1)\\
=&i\int_b^a dx_1\left[A(x_1),\mathcal{O}\right]\\
\mu^2\ \text{term}=&\frac{(i)^2}{2}\mathcal{P}^*\left(\int_b^a dx_1\int_b^a dx_2 A(x_1)A(x_2)\right)\mathcal{O}+(i)\int_b^a dx_1 A(x_1)\mathcal{O}(-i)\int_b^a dx_1 A(x_1),\\
&+\mathcal{O}\frac{(-i)^2}{2}\mathcal{P}^*\left(\int_b^a dx_1\int_b^a dx_2 A(x_1)A(x_2)\right)\\
=&\frac{(i)^2}{2}\mathcal{P}^*\left(\int_b^a dx_1\int_b^a dx_2 A(x_1)A(x_2)\right)\mathcal{O}-\frac{(i)^2}{2}\mathcal{P}^*\left(\int_b^a dx_1\int_b^a dx_2 A(x_1)\mathcal{O}A(x_2)\right)\\
&-\frac{(i)^2}{2}\mathcal{P}\left(\int_b^a dx_1\int_b^a dx_2 A(x_1)\mathcal{O}A(x_2)\right)+\frac{(i)^2}{2}\mathcal{O}\mathcal{P}\left(\int_b^a dx_1\int_b^a dx_2 A(x_1)A(x_2)\right)\\
=&\frac{(i)^2}{2}\mathcal{P}^*\left(\int_b^a dx_1\int_b^a dx_2 A(x_1)\left[A(x_2),\mathcal{O}\right]\right)-\frac{(i)^1}{2}\mathcal{P}\left(\int_b^a dx_1\int_b^a dx_2 \left[A(x_1),\mathcal{O}\right]A(x_2)\right)\\
=&\frac{(i)^2}{2}\tilde{\mathcal{P}}\left(\int_b^a dx_1\int_b^a dx_2\left[A(x_1),\left[A(x_2),\mathcal{O}\right]\right]\right)\\
\cdots&
\end{aligned}
\end{equation}
where $\tilde{\mathcal{P}}$ means the same path order of commutation relations as $\mathcal{P}^*$.

\subsection*{APPENDIX B}
In this appendix we will derive \eqref{eqcom_1}.
For simplicity, we mark $\int_0^{\mu}A(x)dx$ as $\tilde{A}(\mu)$, and have
\begin{equation}
\begin{aligned}
\left[\tilde{A}(\mu),\underline{\gamma}_{ab}\right]=&2i\mu\underline{\hat{T}}_{ab},\\
\left[\tilde{A}(\mu),\underline{\hat{T}}_{ab}\right]=&i\mu\underline{\hat{T}}_{ac}\underline{\gamma}^{cd}\underline{\hat{T}}_{db}.
\end{aligned}
\end{equation}
Then $\left[\tilde{A}(\mu),\underline{\gamma}^{ab}\right]$ is given that
\begin{equation}
\begin{aligned}
\left[\tilde{A}(\mu),\underline{\gamma}^{ab}\right]=&\left[\tilde{A}(\mu),\underline{\gamma}^{ac}\underline{\gamma}_{cd}\underline{\gamma}^{db}\right]\\
=&2\left[\tilde{A}(\mu),\underline{\gamma}^{ab}\right]+\underline{\gamma}^{ac}\left[\tilde{A}(\mu),\underline{\gamma}_{cd}\right]\underline{\gamma}^{db}\\
=&2\left[\tilde{A}(\mu),\underline{\gamma}^{ab}\right]+2i\mu\underline{\gamma}^{ac}\underline{\hat{T}}_{cd}\underline{\gamma}^{db},\\
\left[\tilde{A}(\mu),\underline{\gamma}^{ab}\right]=&-2i\mu\underline{\gamma}^{ac}\underline{\hat{T}}_{cd}\underline{\gamma}^{db}
\end{aligned}
\end{equation}
and $\left[\tilde{A}(\mu),\underline{\hat{T}}_{ac}\underline{\gamma}^{cd}\underline{\hat{T}}_{db}\right]$ is given that
\begin{equation}
\begin{aligned}
\left[\tilde{A}(\mu),\underline{\hat{T}}_{ac}\underline{\gamma}^{cd}\underline{\hat{T}}_{db}\right]=&\left[\tilde{A}(\mu),\underline{\hat{T}}_{ac}\right]\underline{\gamma}^{cd}\underline{\hat{T}}_{db}+\underline{\hat{T}}_{ac}\left[\tilde{A}(\mu),\underline{\gamma}^{cd}\right]\underline{\hat{T}}_{db}+\underline{\hat{T}}_{ac}\underline{\gamma}^{cd}\left[\tilde{A}(\mu),\underline{\hat{T}}_{db}\right]\\
=&i\mu\underline{\hat{T}}_{ac}\underline{\gamma}^{cd}\underline{\hat{T}}_{de}\underline{\gamma}^{ef}\underline{\hat{T}}_{fb}-2i\mu\underline{\hat{T}}_{ac}\underline{\gamma}^{cd}\underline{\hat{T}}_{de}\underline{\gamma}^{ef}\underline{\hat{T}}_{fb}+i\mu\underline{\hat{T}}_{ac}\underline{\gamma}^{cd}\underline{\hat{T}}_{de}\underline{\gamma}^{ef}\underline{\hat{T}}_{fb}\\
=&0.
\end{aligned}
\end{equation}
Finally, we obtain
\begin{equation}
\begin{aligned}
\left[\int_0^{\mu}A(x)dx,\underline{\gamma}^{ab}\right]=&-2i\mu\underline{\gamma}^{ac}\underline{\hat{T}}_{cd}\underline{\gamma}^{db},\\
\left[\int_0^{\mu}A(x)dx,\underline{\hat{T}}_{ac}\underline{\gamma}^{cd}\underline{\hat{T}}_{db}\right]=&0.
\end{aligned}
\end{equation}
\end{appendix}

\end{document}